%% file: index.tex
\documentclass[10pt,conference, twocolumn]{IEEEtran}

\newcommand{\beq}{\begin{eqnarray*}}
\newcommand{\eeq}{\end{eqnarray*}}
\newcommand{\beqn}{\begin{eqnarray}}
\newcommand{\eeqn}{\end{eqnarray}}

\IEEEoverridecommandlockouts
\usepackage{graphics}
\usepackage{epsfig}
\usepackage{times}
\usepackage{amsmath}
\usepackage{amssymb}
\usepackage{verbatim}
\newtheorem {thm}{\bf Theorem} 
\newtheorem {res}{\bf Result} 
\newtheorem {lem}{\bf\noindent Lemma}

\newtheorem {cor}{\bf Corollary}

\newcommand {\SR}[1]{SR_{#1}}
\newcommand {\Rach}{\SR{ach}}

\newcommand {\mymax}[4]{
\begin{align}
{#1}&\textrm{max} \  \ {#2}\label{#4}\\
&\textrm{subject to}   \ {#3}\notag
\end{align}}
\IEEEoverridecommandlockouts

\begin{document}

\title{On the Capacity of One-sided Two user Gaussian Fading Broadcast Channels}

\author{
\authorblockN{Amin Jafarian}
\authorblockA{ECE Department\\
University of Texas at Austin \\
Austin, TX 78712}
\and
\authorblockN{Sriram Vishwanath}
\authorblockA{ECE Department \\
University of Texas at Austin \\
Austin, TX 78712 }
\thanks{A. Jafarian and S. Vishwanath (email: jafarian@ece.utexas.edu; sriram@ece.utexas.edu) are supported in part by National Science Foundation
grants NSF CCF-0448181, NSF CCF-0552741, NSF CNS-0615061, and NSF
CNS-0626903, THECB ARP.}%
}

\maketitle

\input{body2}

\input{appendices}

\bibliographystyle{IEEEtran.bst}
\bibliography{ref}

\end{document}

%% file: body2.tex
\begin{abstract} In this paper, we investigate upper and lower bounds on the capacity of two-user fading broadcast channels where one of the users has a constant (non-fading) channel. We use the Costa entropy power inequality (EPI) along with an optimization framework to derive upper bounds on the sum-capacity and superposition coding to obtain lower bounds on the sum-rate for this channel. For this fading broadcast channel where one channel is constant, we find that the upper and lower bounds meet under special cases, and in general, we show that the achievable sum-rate comes within a constant of the outer bound.
\end{abstract}
\section{Introduction}

\begin{figure}
\centering
\includegraphics[width=3.3in]{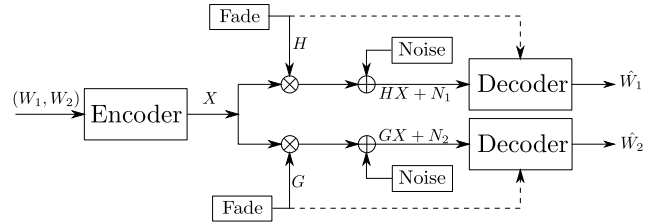}
\caption{Channel Model}
\label{fig}
\end{figure}

The fading broadcast channel is an additive Gaussian noise channel with multiplicative state, as shown in Figure \ref{fig}. The channel state, called the ``fade", is unknown to the transmitter while it is perfectly known at the receiver. Mathematically, a two-user fading broadcast channel can be represented as:

\[
Y_1 = H X + N_1,
\]
and 
\[ Y_2 = G X + N_2,
\]

where $H$ and $G$ are random variables termed the fade states, which are assumed to be discrete-valued in this paper. $X$ and $Y_1,Y_2$ represent the input and the output seen by each receiver (1 and 2) respectively, and $N_i, i \in \{1,2\}$ is additive Gaussian noise assumed to be of unit variance, and the transmit power is assumed to be constrained to be less than $Q$.

There is a large body of work on the Gaussian broadcast channel and its variations \cite{b0, b1}. The capacity region of the Gaussian broadcast channel where $H$ and $G$ are both constant is well known \cite{bergmans}. For the case when $H,G$ are random but are known to transmitter and all receivers, the problem is again solved as it is a set of parallel degraded channels \cite{goldsmith}, \cite{tse}. The Gaussian broadcast channel with additive state has also been well studied \cite{cover}. Finally, the capacity region of MIMO broadcast channel with channel state known to all parties has also been solved \cite{weingarten}.

Recently, the fading broadcast channel has received significant attention \cite{agarwal}, \cite{tuninetti}. The fading broadcast channel is a setting where the realizations of $H,G$ are unknown to the transmitter and are known perfectly to both receivers. Note that the p.m.f. of $H,G$ is assumed to be known to all parties. In \cite{tuninetti}, the authors  determine characterizations on the capacity of a class of these channels. The authors effectively utilize results linking MMSE with entropy in deriving their results. Nevertheless, the capacity region of the general two-user Gaussian fading broadcast channel remains unsolved to date.

In this paper, we find upper and lower bounds on the sum-capacity of this channel using the Costa EPI \cite{EPI:costa} for the case when one channel is constant (i.e. $G=g$ with probability one). We show that our upper and lower bounds meet for non-trivial cases and are in general, a constant distance from one another. By a constant gap, we mean that the difference between them does not grow with transmit power. Note that the major stumbling block in obtaining a characterization on the capacity of these channels is the existence/identification of a suitable entropy power inequality (EPI)\cite{tuninetti}. In this work, we find that the EPI introduced by Costa in \cite{EPI:costa} is a useful tool for obtaining upper bounds on capacity. 

The rest of this paper is organized as follows. The next section discusses the preliminary framework for the outer bound. Further details on the outer bound are discussed in Section \ref{outerbound}.  Section 
\ref{innerbound} compares the outer and inner bound on capacity respectively, showing that the gap between the these bounds is tight in many cases and constant at worst. The paper concludes with Section \ref{conclusion}.

\section{Preliminaries}
\label{background}
We consider a setup in which $G=g$ and $H$ is fading with a finite number of states $\{h_1,h_2,\ldots,h_n\}$. Let $F$ be the vector of fades known at both receivers, i.e. $[H \ g]$. 
As the fade is known at both receivers, we can rewrite the channel model as follows::
\begin{align*}
&Y_1=\left\{\begin{array}{l l}
X+\frac{N_1}{h_1} & \textrm{with probability }p_1\\
 & \\
X+\frac{N_1}{h_2} & \textrm{with probability }{p_2}\\
\vdots\\
X+\frac{N_1}{h_n} & \textrm{with probability }{p_n}
\end{array}\right. ,\\
&Y_2=X+\frac{N_2}{g}
\end{align*}

Throughout this paper we denote sum-rate of two channels by $\SR{}\triangleq R_1+R_2$. Therefore, $\SR{upp}$ is the upper-bound on sum-rate and $\SR{ach}$ denotes the achievable sum-rate.
   
A general upper bound on the sum-rate of fading broadcast channel was obtained by  K$\ddot{\textrm{o}}$rner \& Marton in \cite{outer:km}, which can be simplified to obtain the following expression:
\begin{align}
\SR{} & \le I(X;Y_1|U,F)+I(U;Y_2|F)\notag\\
&=h(Y_1|U,F)-h(Y_1|X,U,F)+h(Y_2|F) \notag\\
&\quad -h(Y_2|U,F)\notag\\
&\le h(Y_1|U,F)-h(Y_2|U,F)+C, \label{ineq}
\end{align}
where $C$ is given by:
\begin{equation}
\label{C_lem}
C=\frac{1}{2}\log\left(Q+\frac{1}{g^2}\right)-\sum_{i=1}^n p_i \frac{1}{2}\log\left( \frac{1}{h_i^2}\right).
\end{equation}

This follows from the following inequalities:
\begin{align*}
h(Y_2|F) &-h(Y_1|X,U,F) \\
&\stackrel{a}{=} h(Y_2|g)-h(Y_1|X,H)\\
&\stackrel{}{=} h(Y_2)-\sum_{i=1}^n p_i h(\frac{N_1}{h_i})\\
&\stackrel{b}{\le}\frac{1}{2}\log\left( 2 \pi e \left( Q+\frac{1}{g^2}\right)\right)-\sum_{i=1}^n p_i \frac{1}{2}\log\left( \frac{2 \pi e}{h_i^2}\right) \\
&=C,
\end{align*}
where (a) follows from the fact that $U \rightarrow X \rightarrow Y_1$ forms a Markov Chain and (b) holds because Gaussian input maximizes $h(Y_2)$.

In the following section, we provide further details on deriving the outer bound by further bounding $h(Y_1|U,F)-h(Y_2|U,F)$.

\section{Outer-bound}
\label{outerbound}

For the remainder of this paper, we use the following notation for convenience:
\[Y_{1i} \triangleq X+\frac{N}{h_i}, \ 
f_i \triangleq h(Y_{2i}|U), \ 
k \triangleq h(Y_{1}|U),\]
for $1\le i\le n$ where $U$ is a random variable, independent of noise $N$.
Also without loss of generality, we assume that $h_1<h_2<...<h_n$.

Now, if either $g>h_n$ or $g<h_1$, the channel is called strongly degraded and its capacity region is known \cite{strongdeg}. Note that, in the strongly degraded case, one channel always dominates the other (across all states). For this work, we determine outer bounds (and achievable rates) when $h_1<g<h_n$.

We use the following lemma, a variation of Costa's EPI \cite{EPI:costa}.
\begin{flushleft}
\begin{lem} 
\label{EPI_u} 
$2^{2h(x+\sqrt{t}N|U)}$ is a concave function with respect to $t$ where $N$ is a Gaussian noise and $U$ is a random variable independent of $N$.
\end{lem}
\end{flushleft}
{\begin{proof}
This is a conditional version of the Costa EPI in [7]. For a full proof, see Appendix \ref{costarev}.
\end{proof}}

Using this lemma,  and our channel model, we can write the following  relationship between the entropy powers:
\begin{equation}
\label{z_upp}
2^{2k} \ge \sum_{i=1}^n \alpha_i 2^{2f_i},
\end{equation} 
for all $\alpha_i$s that satisfy the following conditions:
\begin{equation}
\begin{split}
\label{alpha_cond}
&\alpha_i \ge 0 \ \ \forall i, \ \ \sum_{i=1}^n \alpha_i = 1 , \ \ \sum_{i=1}^n \frac{\alpha_i}{h_i^2}=\frac{1}{g^2}.
\end{split}
\end{equation}

Combining Equation (\ref{z_upp}) with Equation (\ref{ineq}), we have the following optimization problem as an upper bound on the sum-capacity of the channel:
\mymax{}{\sum_{i=1}^{n}p_if_i-k+C,}{
2^{2k} \ge \sum_{i=1}^n \alpha_i 2^{2f_i}
}{}
where $\alpha_i\textrm{'s satisfy }(\ref{alpha_cond})$.
 As $C$ is a constant, this reduces to solving:
 
\mymax{}{\sum_{i=1}^{n}p_if_i-k.}{
2^{2k} \ge \sum_{i=1}^n \alpha_i 2^{2f_i}}{s_rate} 

Note that due to the power constraint for each $f_i$,  we have:
\begin{equation}
\label{fi_const}
\frac{1}{2}\log\left(\frac{2 \pi e}{h_i^2}\right)\le f_i \le \frac{1}{2}\log\left( 2 \pi e \left( Q+\frac{1}{h_i^2}\right)\right).
\end{equation} 
From (\ref{fi_const}), it follows that we can upper bound the sum-rate in Equation (\ref{s_rate}) as:
\mymax{D = \ }{ \sum_{i=1}^{n}p_if_i-\frac{1}{2}\log\left(\sum_{i=1}^n \alpha_i 2^{2f_i}\right).}{
\begin{array}[t]{c}
\frac{1}{2}\log\left(\frac{2 \pi e}{h_i^2}\right)\le f_i \le \frac{1}{2}\log\left( 2 \pi e \left( Q+\frac{1}{h_i^2}\right)\right)
\end{array}
}{conc_upp}

So from the above discussion and (\ref{C_lem}), we have $\SR{} \le \SR{upp}$ where $\SR{upp}$ is defined as the following:
\begin{equation}
\label{bounding}
\SR{upp}=D+\frac{1}{2}\log\left( Q +\frac{1}{g^2}\right)-\sum_{i=1}^n \frac{p_i}{2}\log\left(\frac{1}{h_i^2}\right).
\end{equation}
It is easy to show that the objective function in (\ref{conc_upp}) is in fact jointly strictly concave in all the variables $f_i$ \cite{boyd}\footnote{The objective in our setting is an objective of a geometric program}. 
To solve it, we initially ignore the linear boundary constraints in (\ref{fi_const}) and differentiate to determine the maximizing point of the jointly concave function. Subsequently, we check the maximizing point to ensure it meets the boundary constraints imposed by (\ref{fi_const}). This gives us the following equations  that must be satisfied by the optimizing $\alpha_r$ and $x_r$, with the $\alpha_r$ belonging to the feasible set given by (\ref{alpha_cond}):
\begin{equation}
\label{f_cond}
p_r=\frac{\alpha_r 2^{2f_r}}{\sum_{i=1}^n \alpha_i 2^{2f_i}} \qquad \forall r\in \{1,2,\dots,n\}.
\end{equation}

The main effort at this stage is to show that feasible $f_r$ and $\alpha_r$ exist, and to determine the maximizing value for (\ref{conc_upp}). To do this, we define the following function:
\begin{equation}
\label{main_equ}
T(x)=\sum_{i=1}^n \frac{p_i}{x+\frac{1}{h_i^2}}-\frac{1}{x+\frac{1}{g^2}}.
\end{equation}

The following lemma characterizes the roots of this function.
{\flushleft\begin{lem}
\label{outside_sol}
The function $T(x)$ as defined in Equation (\ref{main_equ}) always has exactly one solution  outside the interval $[-\frac{1}{h_1^2},-\frac{1}{h_n^2}]$. We define $x^*$ to be this root.
\end{lem}}
\begin{proof}
The degree of polynomial in the numerator of $T(x)$ is $n-1$, and $T(x)$ has exactly $n-2$ roots in $[-\frac{1}{h_1^2},-\frac{1}{h_n^2}]$, which means that it has one root outside this interval. The complete proof is given in Appendix \ref{one_x_sol}.
\end{proof}
Now depending on the value of  $x^*$, we obtain the following three cases:

\begin{equation}
\begin{split}
\label{cases}
\textrm{\bf{Case 1:}} \  x^* \in A_1 =& \ [0,Q].\\
\textrm{\bf{Case 2:}} \ x^* \in A_2 =& \ [-\infty,-Q-\frac{2}{h_1^2}]\cup[-\frac{1}{h_n^2},0] \\
& \ \cup[Q,+\infty].\\
\textrm{\bf{Case 3:}} \ x^* \in A_3=& \  [-Q-\frac{2}{h_1^2},-\frac{1}{h_1^2}]. \qquad \qquad \qquad \ 
\end{split}
\end{equation}

We address each of the above cases separately.
The following lemma gives a characterization of our maximization problem if Case 1 holds.
{\flushleft \begin{res}
\label{case_one}
If $x^* \in A_1$, the solution for (\ref{conc_upp})  is given by:
\begin{equation}
\label{D_case1}
D=\sum_{i=1}^n \frac{p_i}{2}\log \left( x^*+\frac{1}{h_i^2}\right)- \frac{1}{2}\log\left( x^*+\frac{1}{g^2}\right).
\end{equation}
\end{res}}
\begin{proof}
 We show that a feasible $\alpha_i$ and $f_i$ can be found for this case. Consider:
\begin{equation}
\label{lem_alpha}
\alpha_i=p_i \left[\frac{x^*+\frac{1}{g^2}}{x^*+\frac{1}{{h_i^2}}}\right].
\end{equation}
This assignment meets the  constraints in (\ref{alpha_cond}). Simultaneously,  (\ref{f_cond}) will be satisfied by letting 
$f_i=\frac{1}{2}\log\left( 2 \pi e \left( x^*+\frac{1}{h_i^2}\right)\right).$
Note that, as already mentioned, problem (\ref{conc_upp}) is strictly concave, so if there exists a solution satisfying conditions  (\ref{f_cond}), it is the only point which maximizes (\ref{conc_upp}). 

We can compute $D$ in this case as:
\begin{align}
D&= \sum_{i=1}^n \frac{p_i}{2}\log\left( 2 \pi e \left( x^*+\frac{1}{h_i^2}\right)\right)\notag \\
&\qquad - \frac{1}{2}\log\left(2 \pi e\left(\sum_{i=1}^n \alpha_i\left(x^*+\frac{1}{h_i^2}\right)\right)\right)\notag\\
&=\sum_{i=1}^n \frac{p_i}{2}\log\left( x^*+\frac{1}{h_i^2}\right)- \frac{1}{2}\log\left( x^*+\frac{1}{g^2}\right),\notag
\end{align}
where the last line follows from properties of $\alpha_i$ given in (\ref{alpha_cond}).
\end{proof}

Now consider Case 2. In this case, we find that $D$ is maximized on the boundary of (\ref{fi_const}). The following result summarizes our result in this case:
{\flushleft \begin{res}
\label{case_two}
Maximum of (\ref{conc_upp}) is attained on the boundary of optimization problem (given by (\ref{conc_upp})) if $x^*$ occurs  in one of the following intervals \footnote{Note that $B_1\cup B_2\cup B_3=A_2$}:
\begin{itemize}
\item $B_1=[-\infty,-Q-\frac{2}{h_1^2}]$
\item $B_2=[-\frac{1}{h_n^2},0]$
\item $B_3=[Q,+\infty]$
\end{itemize}
\end{res}}
\begin{proof}
Let $\alpha_i$'s be as above in (\ref{lem_alpha}). This assignment still satisfies (\ref{alpha_cond}) for all the intervals above. 

First consider the case when $x^* \in B_1$.
From the strict concavity of (\ref{conc_upp}), it is easy to show that $D$ will be maximized by setting $\hat{f_i}$ equal to:
\begin{equation}
\label{boundary_cond}
\hat{f_i}=\frac{1}{2}\log\left( 2 \pi e \left( -x^*-\frac{1}{h_i^2}\right)\right),
\end{equation}
as this choice satisfies the conditions given in (\ref{f_cond}). In addition,  $-x^*-\frac{1}{h_i^2}>Q+\frac{1}{h_i^2}$ for every $i$ for  $ x^* \in B_1$, which implies that $\hat{f_i}>\frac{1}{2}\log\left( 2 \pi e \left( Q+\frac{1}{h_i^2}\right)\right)$. Using concavity of (\ref{conc_upp}) again, it is easy to see that in this case $f_i=\frac{1}{2}\log\left( 2 \pi e \left( Q+\frac{1}{h_i^2}\right)\right)$, which meets with the boundary of the feasibility region in (\ref{fi_const}).

If $x^* \in B_2$ or $B_3$, the same proof can be repeated to show the solution is on the boundary of the feasibility region of (\ref{conc_upp}).

When $x^* \in B_1$ or $B_3$, the maximum of problem (8) can be computed as:
\begin{equation}
\label{D_case2_13}
D=\sum_{i=1}^n \frac{p_i}{2}\log\left( Q+\frac{1}{h_i^2}\right)- \frac{1}{2}\log\left( Q+\frac{1}{g^2}\right),
\end{equation}
and if $x^* \in B_2$, we get the following value:
 \begin{equation}
\label{D_case2_2}
D=\sum_{i=1}^n \frac{p_i}{2}\log\left(\frac{1}{h_i^2}\right)- \frac{1}{2}\log\left(\frac{1}{g^2}\right).
\end{equation}

\end{proof}

Finally, consider Case 3, i.e when $x^* \in A_3$.

{\flushleft \begin{res}
\label{case_three}
If $x^* \in A_3$, $D$ is bounded by the following expresion:
\begin{equation}
\label{D_case3}
D\le \sum_{i=1}^{n}\frac{p_i}{2}\log\left( -x^*-\frac{1}{h_i^2}\right)-\frac{1}{2}\log\left( -x^*-\frac{1}{g^2}\right).
\end{equation}
\end{res}}
\begin{proof}
Using $\alpha_i$ as in (\ref{lem_alpha}) again leads to a feasible choice of parameters. To get the upper bound here, we drop the conditions on $f_i$'s given by (\ref{fi_const}).
\end{proof}

With these last three results, we conclude the discussion on an upper bound on the sum-rate capacity of this channel. In the next section, we provide an achievable scheme for this channel and show that it lies, in the worst case, a constant gap away from the outer bounds obtained.

\section{Inner-bound}
\label{innerbound}
In this section we address the gap between the outer bound found in Section \ref{outerbound} and the achievable scheme that uses superposition coding. It is easy to show that superposition yields the following achievable sum-rate for this channel \cite{cover}:
\begin{equation}
\label{R_ach}
\Rach=\max_{0\le \beta\le 1} \frac{1}{2}\log\left( 1+\beta g^2 Q\right)+\sum_{i=1}^n{\frac{p_i}{2}\log\left(\frac{1+h_i^2 Q}{1+\beta h_i^2 Q}\right)}.
\end{equation}

The following lemma solves the above maximization problem:
{\flushleft\begin{lem}
\label{R_ach_lem}
The optimization problem in (\ref{R_ach}) attains its maximum at either $\beta=0$ or $\beta=1$. 
\end{lem}}
\begin{proof}
In order to prove this lemma it is sufficient to show that objective in (\ref{R_ach}) is either convex or strictly increasing/decreasing. It then follows from convex optimization arguments that the solution lies on the boundary of the set $0\le \beta \le 1$ \cite{boyd}. Let us call this function $\SR{ach}(\beta)$.

The proof follows immediately if the $\SR{ach}(\beta)$ is strictly increasing/decreasing. Thus, let us  assume that the function is neither strictly increasing nor decreasing. Given that it is differentiable, it follows that the derivative given by:
\[
\frac{1}{\frac{1}{g^2}+\beta Q} - \sum_{i=1}^n \frac{p_i}{\frac{1}{h_i^2}+\beta Q}=0
\]
has a solution in the interior of $0\le \beta\le 1$. Call this solution $\beta^*$. Then we have the following:
\begin{align*}
\frac{1}{\frac{1}{g^2}+\beta^* Q} &= \sum_{i=1}^n \frac{p_i}{\frac{1}{h_i^2}+\beta^* Q} 
\stackrel{(a)}{\le} \sqrt{\sum_{i=1}^n \frac{p_i}{(\frac{1}{h_i^2}+\beta^* Q)^2}},
\end{align*}
where (a) follows from Jensen's inequality \cite{bcover}. Note that this implies that the $\SR{arch}(\beta)$ is convex, as 
\[
-\frac{1}{(\frac{1}{g^2}+\beta^* Q)^2} + \sum_{i=1}^n \frac{p_i}{(\frac{1}{h_i^2}+\beta^* Q)^2} \ge 0, 
\]
which is the second derivative of $\SR{ach}(\beta)$ with respect to $\beta$. This concludes the proof.
\end{proof}
{\flushleft\begin{cor}
\label{R_cor}
$\Rach$ is bounded below by $\frac{1}{2}\log\left(1+g^2 Q\right)$.
\end{cor}}
In the next theorem, we compare the achievable rates with the outer bounds derived in Section \ref{outerbound}.

{\flushleft\begin{thm}
 Consider the three cases as defined in Equation (\ref{cases}). When the upper bound belongs to Cases 1 and 3 ( i.e., the upper bound corresponds to Equations (\ref{D_case1}) and (\ref{D_case3})), there is a computable constant gap between the achievable rates and the upper bound. When the upper bound belongs to Case 2 (and evaluates to values in Equations (\ref{D_case2_13}) or (\ref{D_case2_2})), the achievable rate and upper bound are equal.
\end{thm}}
\begin{proof} The proof first considers Case 1  in (\ref{cases}). The gap between the lower and upper bounds can be written as: 

\begin{align*}
\textrm{Gap}&=\SR{upp}-\Rach\\ 
&{\le} D+\frac{1}{2}\log\left( Q+\frac{1}{g^2}\right)-\sum_{i=1}^n p_i \frac{1}{2}\log\left(\frac{1}{h_i^2}\right)\\
& \qquad -\frac{1}{2}\log\left(1+g^2 Q\right)\\
&\stackrel{(a)}{=}\sum_{i=1}^n \frac{p_i}{2}\log\left(h_i^2 x^*+1\right)-\frac{1}{2}\log\left(g^2x^*+1\right),
\end{align*}
where (a) follows from Equation (\ref{D_case1}).
Note that this gap is only a function of the channel parameters, as $x^*$, the root of $T(x)$, is only a function of channel parameters and not of $Q$. Thus, as $Q$ increases, this gap does not increase.

The gap in Case 3 can be written in a similar fashion:
\begin{align*}
\textrm{Gap}&=\SR{upp}-\Rach\\ 
&{\le} D+\frac{1}{2}\log \left( Q+\frac{1}{g^2}\right)-\sum_{i=1}^n p_i \frac{1}{2}\log\left(\frac{1}{h_i^2}\right)\\
& \qquad -\frac{1}{2}\log\left(1+g^2 Q\right)\\
&\stackrel{(b)}{=}\sum_{i=1}^n \frac{p_i}{2}\log\left(-h_i^2 x^*-1\right)-\frac{1}{2}\log\left(-g^2x^*-1\right),
\end{align*}
where (b) follows from Equation (\ref{D_case3}).
For the same reasons as Case 1 above, this gap is a constant as well and does not increase with transmit power.

Finally, we analyze the setting when the upper bound is in Case 2 above (\ref{cases}).
In this case we prove that the achievable sum-rate and upper bounds meet. Consider the following function:
\begin{align*}
g(x)&=\Rach |_{Q=0}-\Rach |_{Q=1}\\
&=\sum_{i=1}^n \frac{p_i}{2}\log(1+h_i^2 x)-\frac{1}{2}\log(1+g^2 x).
\end{align*}

Note that the derivative of $g(x)$ with respect to $x$ is equal to $T(x)$. 
One can check that:
\begin{equation}
\label{limit}
T(x) \rightarrow +\infty \textrm{ as } x \rightarrow \left(-\frac{1}{h_n^2}\right)^+
\end{equation}

As $A_2=B_1\cup B_2\cup B_3$, the union of three disjoint intervals, we discuss each interval $B_i, 1\le i\le3$ separately.

\noindent{\bf Setting 1}{ ($x^* \in B_1$):} 
It follows from Lemma \ref{outside_sol} and Equation (\ref{limit}) that $T(x)>0$ for all $x>-\frac{1}{h_n^2}$. So $g(x)$ is increasing for $x \ge 0$ and $g(Q)>g(0)=0$. 
Optimality of achievable scheme in this setting can be obtained from Result \ref{case_two}, Lemma  \ref{R_ach_lem}  and Equation (\ref{bounding}) as:
\begin{align*}
\textrm{Gap}&=\SR{upp}-\Rach\\
&\le \sum_{i=1}^n \frac{p_i}{2}\log\left( Q+\frac{1}{h_i^2}\right)-\frac{1}{2}\log\left(Q+\frac{1}{g^2}\right)\\
&\qquad  +C-\sum_{i=1}^n \frac{p_i}{2}\log\left(1+h_i^2 Q\right)
=0
\end{align*}

\noindent{\bf Setting 2}{ ($x^* \in B_2$):} 
Similar to Setting 1, it follows that $T(x)<0$ for all $x>x^*$. So, $g(x)$ is decreasing for all $x>0$ which means $g(Q)<g(0)=0$. Thus the gap in this case is:
\begin{align*}
\textrm{Gap}&=\SR{upp}-\Rach\\
&\le \sum_{i=1}^n \frac{p_i}{2}\log\left(\frac{1}{h_i^2}\right)-\frac{1}{2}\log\left(\frac{1}{g^2}\right)+C\\
&\qquad-\frac{1}{2}\log\left(1+g^2 Q\right)
=0,
\end{align*}

\noindent{\bf Setting 3 }{($x^* \in B_3$):} 
This case is similar to the case where $x^* \in B_1$. Again we find that $T(x)>0$ for all $x<x^*$ which includes all $x<Q$. Therefore, $g(x)$ is increasing for all \protect{$0<x<Q$}. Following the same lines as Setting 1,  we get that the upper and lower bounds meet in this setting as well.

\end{proof}

Intuition behind these results: Assuming that Gaussian inputs are optimal, 
we wish to determine the optimal input power split and $\alpha_i$'s satisfying the three conditions given in (\ref{alpha_cond}). 
These $\alpha_i$'s can be computed (as given by (\ref{lem_alpha})) from Equation (\ref{f_cond}) such that they satisfy the third condition. 
The key to the definition of the function T(x) is 
to find an optimal power $x^*$ for Gaussian inputs such that $\alpha_i$'s satisfy the second condition. 
Finally, Lemma \ref{outside_sol} provides the first condition for these $\alpha_i$'s.
 


\section{Conclusion}
\label{conclusion}
In this work, we provide upper and lower bounds on the sum-capacity of a non-degraded broadcast channel - the one-sided two-user Gaussian fading broadcast channel. Using a modified version of Costa's EPI, we derive an upper bound for this channel, and compare it with an achievable scheme that uses superposition coding. We show that the gap between these two does not grow with transmit power and is tight in many cases.


%% file: appendices.tex
\appendix
\subsection{Proof of Lemma \ref{EPI_u}}
\label{costarev}
Let $N(W)=2^{2h(W)}$ and define the function $g_X(t)\triangleq \frac{N(X+\sqrt{t}Z|U)}{N(X|U)}$ and very similarly $f_X(t)\triangleq \frac{N(X+\sqrt{t}Z)}{N(X)}$. We can simplify $g_X(t)$ as the following:
\begin{align*}
g_X(t)&=2^{\frac{2}{n}I(X+\sqrt{t}Z;Z|U)}\stackrel{a}{=}2^{\frac{2}{n}I(X+\sqrt{t}Z;Z)}\\
&=\frac{N(X+\sqrt(t)Z)}{N(X)}=f_X(t),
\end{align*}
where (a) follows from the independence of $Z$ and $U$.
In \cite{EPI:OR} it is shown that $f''_X(t)<0$ and consequently $g''_X(t)<0$. Because $N(X|U)$ is not a function of $t$, we infer that: \[\frac{d^2}{dt^2}N(X+\sqrt{t}Z|U)<0,\] which completes the proof.

\subsection{Proof of Theorem \ref{outside_sol}}
\label{one_x_sol}
Combine terms of $f(x)$ into one ratio of polynomials, $f(x)=\frac{l(x)}{m(x)}$. Note that $l(x)$ has degree exactly $n-1$, so $f(x)$ has at most $n-1$ roots.
Assume the following order for the fades:
\[\frac{1}{h_1^2}\ge \frac{1}{h_2^2} \ge ...\ge \frac{1}{h_k^2} \ge \frac{1}{g^2} \ge \frac{1}{h_{k+1}^2} \ge ... \ge \frac{1}{h_n^2}\]
One can check that:
\begin{itemize}
\item[(a).] $f(x) \rightarrow -\infty$ if $x \rightarrow -{\frac{1}{h_i^2}}^-$ for every $1 \le i \le n$.
\item[(b).] $f(x) \rightarrow +\infty$ if $x \rightarrow -{\frac{1}{h_i^2}}^+$ for every $1 \le i \le n$.
\end{itemize}

From (a) and (b) we can see that $f(x)$ has odd number\footnote{Therefore it has at least one root in this interval} of roots \footnote{Including multiplicity of a root} in $[-\frac{1}{h_i^2},-\frac{1}{h_{i+1}^2}]$ for every $1 \le i \le n-1$ except $i=k$; which totals up to $n-2$ roots. Next, we show that there do not exist any other roots between $-\frac{1}{h_k^2}$ and $-\frac{1}{h_{k+1}^2}$.
Note that:
\begin{itemize}
\item[(1).]  $f(x) \rightarrow +\infty$ if $x \rightarrow -{\frac{1}{g^2}}^-$.
\item[(2).]  $f(x) \rightarrow -\infty$ if $x \rightarrow -{\frac{1}{g^2}}^+$.
\end{itemize}

(1),(2) and (a),(b)  together imply that $f(x)$ has even number of roots in $(-\frac{1}{h_k^2},-\frac{1}{h_{k+1}^2})$, but as we know that it can have at most $(n-1)-(n-2)=1$ root \footnote{Which is exactly one root because we find the other $n-2$ roots and by dividing we leave with a linear equation which gives the other root}. Thus, $f(x)$ has no roots within this interval and one root outside $[-\frac{1}{h_1^2},-\frac{1}{h_n^2}]$.